\documentclass[runningheads]{llncs}
\usepackage{graphicx}
\usepackage{graphics}
\usepackage{xcolor}
\usepackage{caption}
\usepackage{subcaption}
\usepackage{fancyhdr}
\begin{document}
\title{Learning while Acquisition: Towards Active Learning Framework for Beamforming in Ultrasound Imaging\thanks{The authors would like to acknowledge the funding from the Department of Science and Technology - Science and Engineering Research Board (DST-SERB (ECR/2018/001746)) and the support from NVIDIA Corporation for the donation of CLARA AGX Developer Kit through the Academic Hardware Grant program.}}
\titlerunning{Towards Active Learning Framework for Ultrasound Beamforming}
%
\author{Mayank Katare \and  
Mahesh Raveendranatha Panicker \and 
Madhavanunni  A. N. \and 
 Gayathri Malamal} 
\authorrunning{Mayank et al.}
%
\institute{Department of Electrical Engineering and Center for Computational Imaging \\ 
Indian Institute of Technology Palakkad \\
\email{Email:mahesh@iitpkd.ac.in}}

\maketitle              
\begin{abstract}
In the recent past, there have been many efforts to accelerate adaptive beamforming for ultrasound (US) imaging using neural networks (NNs). However, most of these efforts are based on static models that are trained to learn a single adaptive beamforming approach (e.g., minimum variance distortionless response (MVDR)) assuming that they result in the best image quality. Moreover, the training of such NNs is initiated only after acquiring a large set of data that consumes several gigabytes (GBs) of storage space. In this study, an active learning framework for beamforming is described for the first time in the context of NNs. The best quality image chosen by the user serves as the ground truth for the proposed technique, which trains the NN concurrently with data acquisition. On an average, the active learning approach takes 0.5 seconds to complete a single iteration of training.   

\end{abstract}
\section{Introduction}
Medical ultrasound (US) has been a popular non-invasive diagnostic imaging tool with widespread applications in cardiac, abdominal, fetal, musculoskeletal, and breast imaging \cite{travis1988ultrasonic}. Typically US imaging employs a pulse-echo technique where a piezoelectric transducer array transmits acoustic pulses into the tissue and receives the reflections, which are beamformed to produce an output image.
\begin{figure}[tb]
\includegraphics[width=\textwidth]{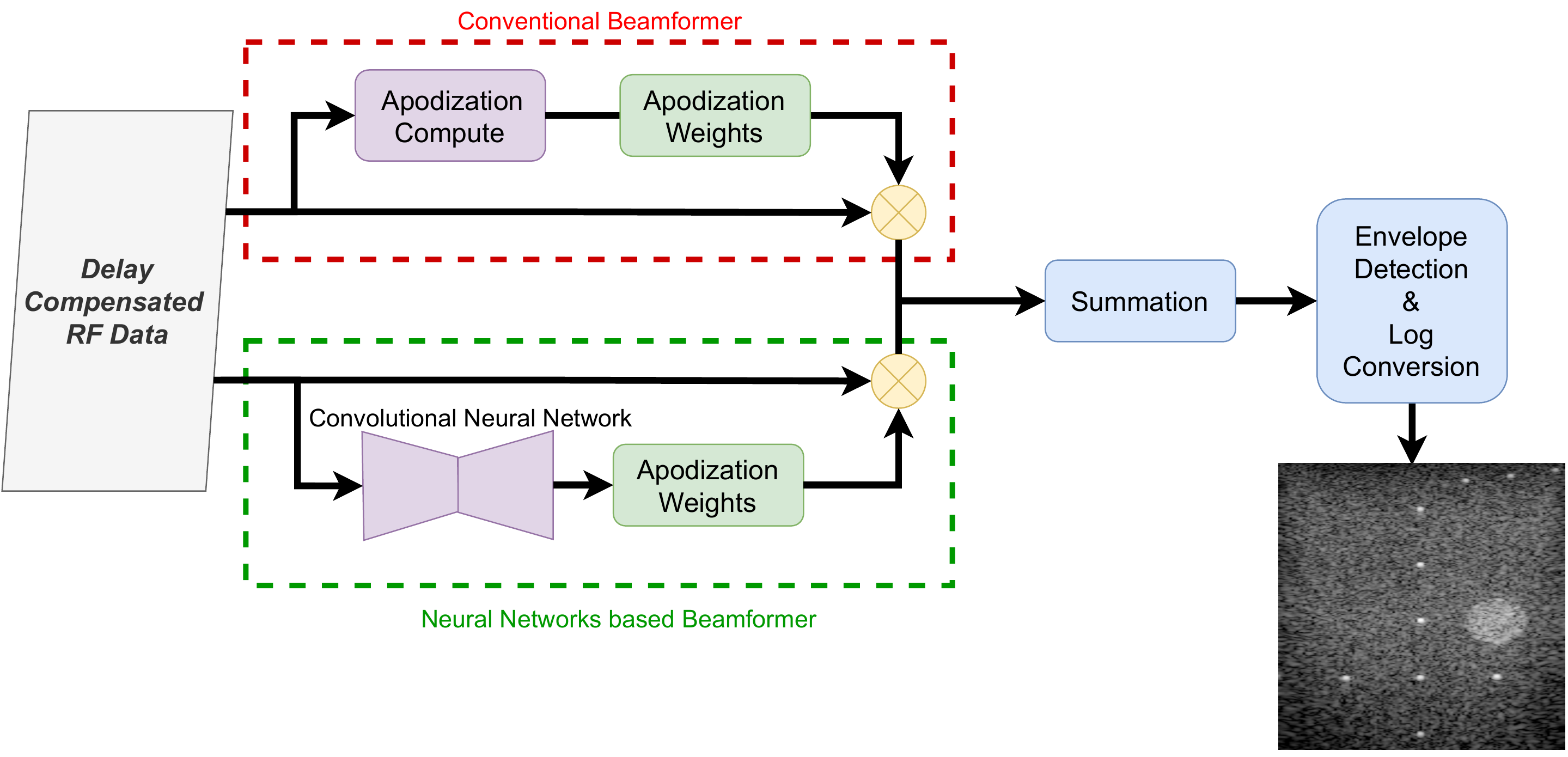}
\caption{Workflow from delay compensated RF data to Brightness (B)-mode image in conventional (upper pipeline (red)) and convolutional neural network (CNN) based  (Bottom pipeline (green)) beamformers. In the CNN based approach, the main objective is to replace and accelerate the computationally intensive estimation of apodization weights. The process involves, the delay compensated RF data as input (of size $2400 \times 128 \times 128$), multiplication of the estimated apodization weights (of size $2400 \times 128 \times 128$) with the delay compensated data and channel wise sum to get the beamformed data (of size $2400 \times 128$). This is further envelope detected and Log compressed to the desired dynamic range.} \label{fig1}
\end{figure}
During beamforming, the raw data acquired from the piezoelectric transducer array is space mapped and converted to form an image as shown in the upper pipeline (red dotted rectangle) of Fig. \ref{fig1}. Commercial US systems use low-complex, delay-and-sum (DAS) \cite{GarciaDAS2021} beamforming to enable real-time reconstruction. DAS applies pre-determined delays and geometry-driven weights to the individual transducer (channel) signals, and sums up the individual contributions to yield the final beamformed signal. Advanced beamforming algorithms like filtered delay multiply and sum (F-DMAS) \cite{FDMAS2015}, minimum variance distortionless response (MVDR) \cite{MVDR2009} and generalized coherence factor (GCF) \cite{li2003adaptive} based beamformers that provide considerable improvement in the image quality over DAS have also been proposed. However, these beamforming schemes are computationally intensive \cite{luijten2020adaptive} which limits the real-time implementation and consequently the adaptability for clinical usage. 

Accelerating computationally complex beamforming algorithms with deep neural networks has been gaining acceptance in the recent past \cite{luijten2020adaptive,eldar_dlus,mia_pwdl,mathews2021towards}. A fully connected neural network \cite{luijten2020adaptive} and a fully connected convolutional network with an identical number of samples along the depth and lateral directions throughout the network \cite{mathews2021towards} have been developed. Recently, in \cite{mia_pwdl}, a self-supervised learning based approach has also been proposed. A general framework for beamforming acceleration with convolutional neural networks is shown in the bottom pipeline (green dotted rectangle) of Fig. \ref{fig1}. However, the NN based approaches in the literature have two concerns: 
1) the bias in the assumption that the best image is always from a specific beamforming scheme (e.g. MVDR), and 2) the training is initiated post acquiring a significant amount of radio-frequency (RF) data. 

In this work, a novel active learning framework is proposed for US beamforming. To the best of our knowledge, this is a first of its kind in the literature. The proposed framework aims at: 
\begin{enumerate}
    \item Beamforming-agnostic learning by training the NN based on the ground truth selected by the user and not specific to any beamforming.
    \item Reduced data storage compared to the conventional training methods. 
    \item Integration of data acquisition and the NN training into a single workflow. 
\end{enumerate}

\begin{figure}[t]
\centering
\begin{subfigure}[t]{0.8\textwidth}
\centering
\includegraphics[width=\textwidth]{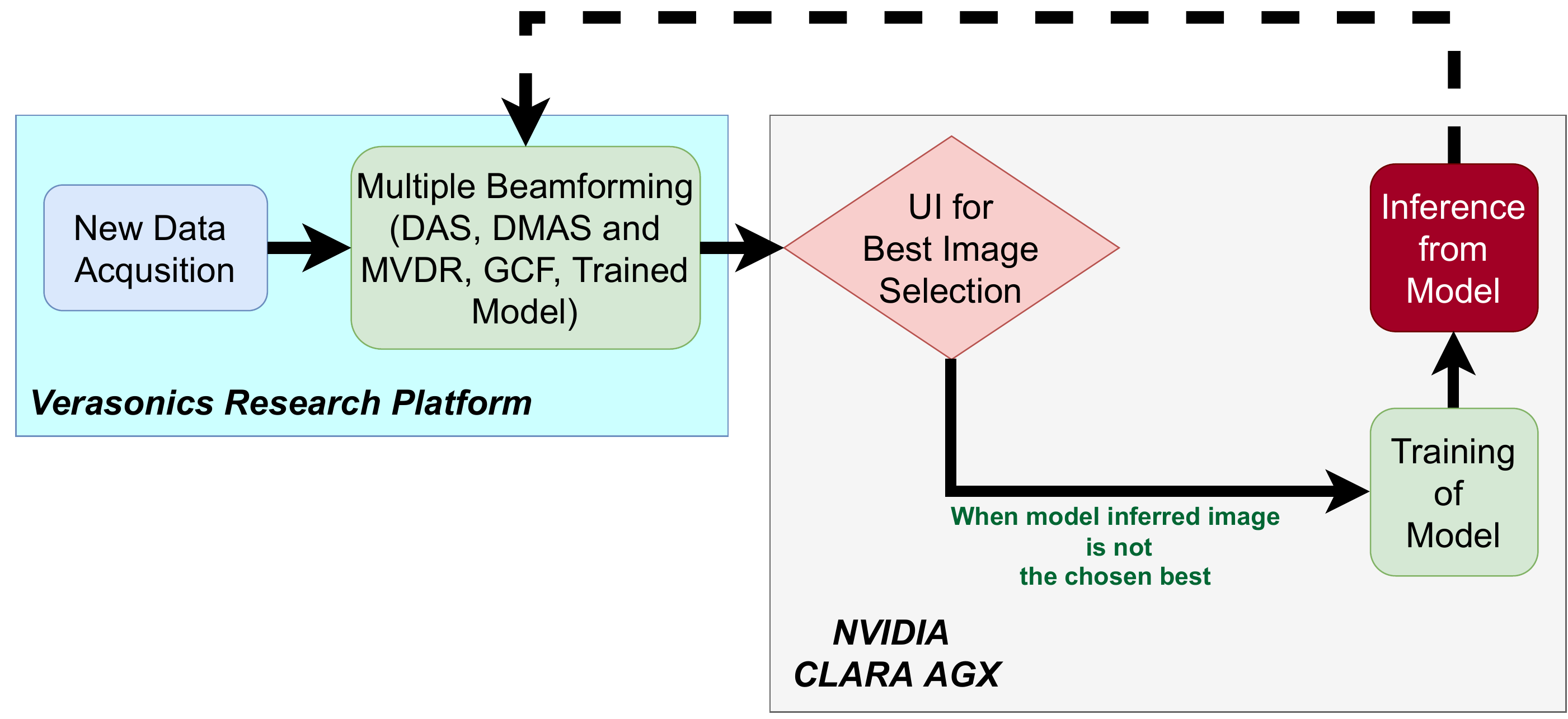}
\caption{ }
\label{fig2a}
\end{subfigure}
\vfill
\begin{subfigure}[b]{0.8\textwidth}
\centering
\includegraphics[width=\textwidth]{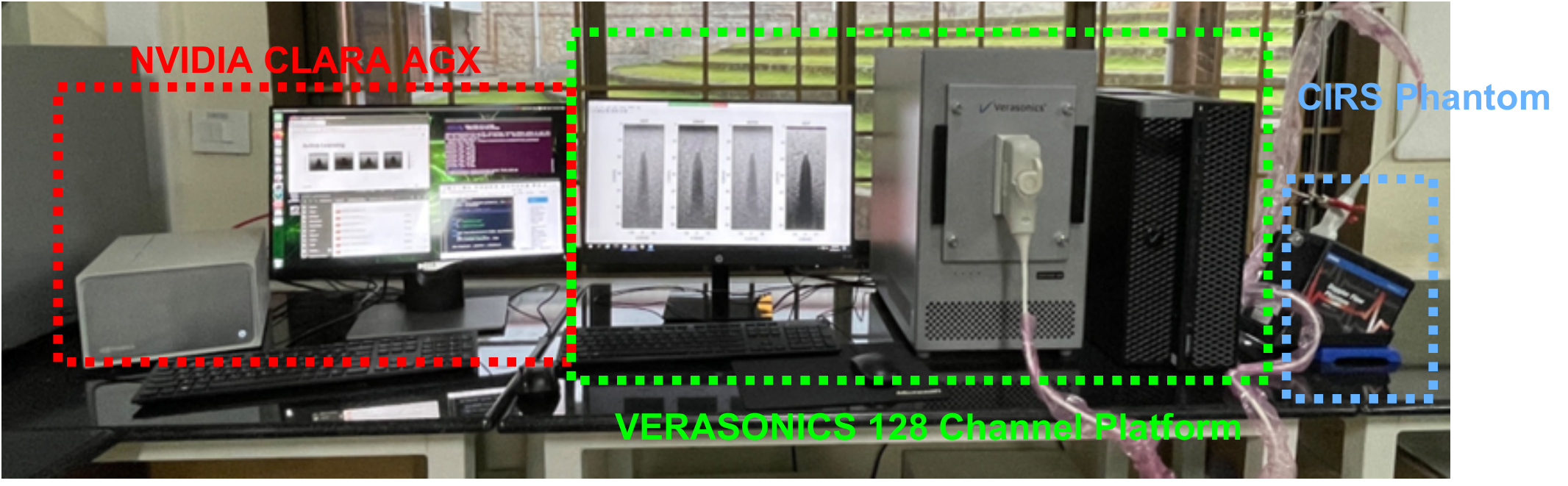}
\caption{ }
\label{fig2b}
\end{subfigure}
\caption{(a) Proposed active learning framework for beamforming. The important steps are generation of images based on DAS, DMAS, MVDR, GCF and proposed model based beamforming which is shown to the user for the selection of the best image, followed by the training of the model if the user selection is different from the model based beamforming image. (b) Data acquisition setup for the proposed active learning frame work. The data acquisition is done using the Verasonics 128 channel research platform and the UI and the training of the model is done on the NVIDIA CLARA AGX Developer Kit. The CIRS Model 769 flow phantom employed for one of the acquisitions is also shown.} 
\label{fig2}
\end{figure}

\section{Active Learning for US Beamforming}
The proposed framework is as shown in Fig. \ref{fig2a} and the experimental setup is illustrated in Fig. \ref{fig2b}. The training data is acquired with the Verasonics Vantage 128 research US platform using 128 channel L11-5v linear array transducer with a center frequency of 7.6 MHz for non-steered plane wave transmission. The active learning is implemented as a lightweight U-Net based multi-scale NN to accelerate beamforming on a NVIDIA CLARA AGX Developer Kit. 

\begin{figure}[t]
\centering
\includegraphics[width=\textwidth]{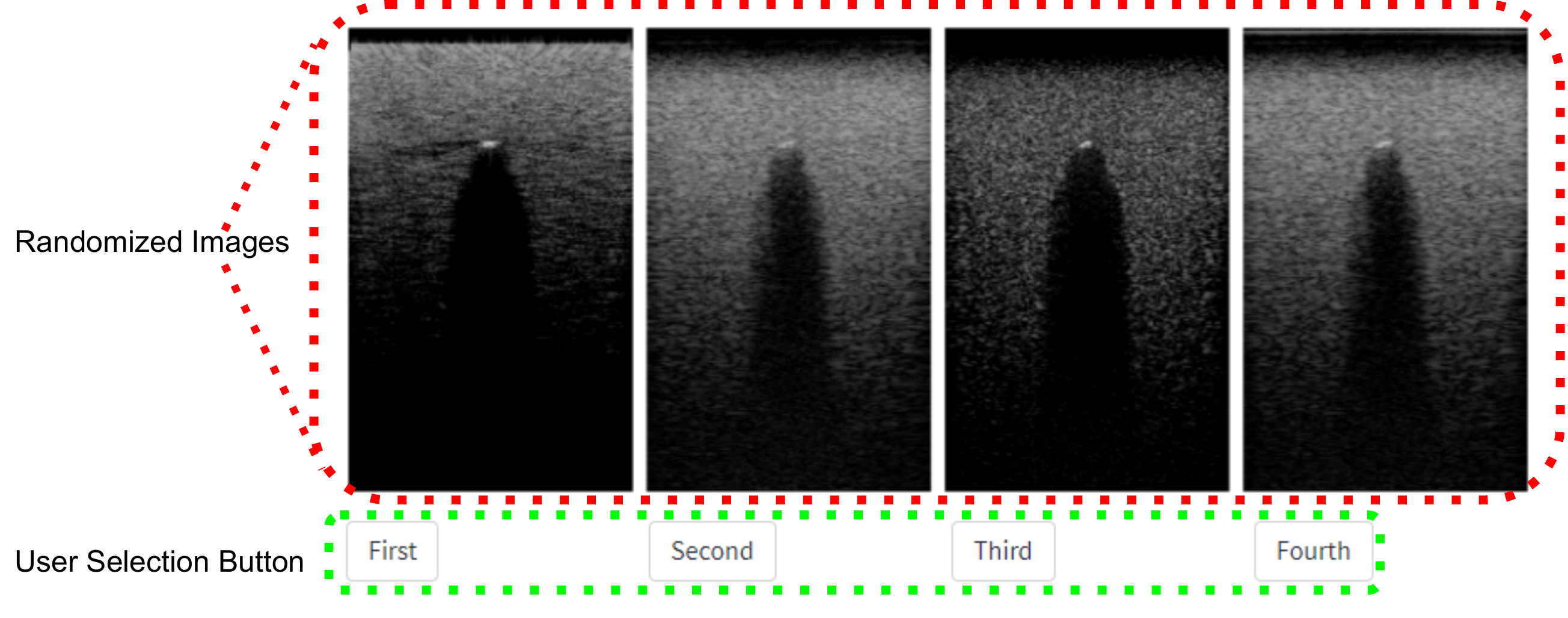}
\caption{GUI of Active Learning Framework. The images generated by DAS, DMAS, MVDR and GCF are randomized and provided to the user. In this case, the image generated by the model is not shown.} \label{fig3}
\end{figure}

\begin{figure}[b]
\centering
\includegraphics[width=0.8\textwidth]{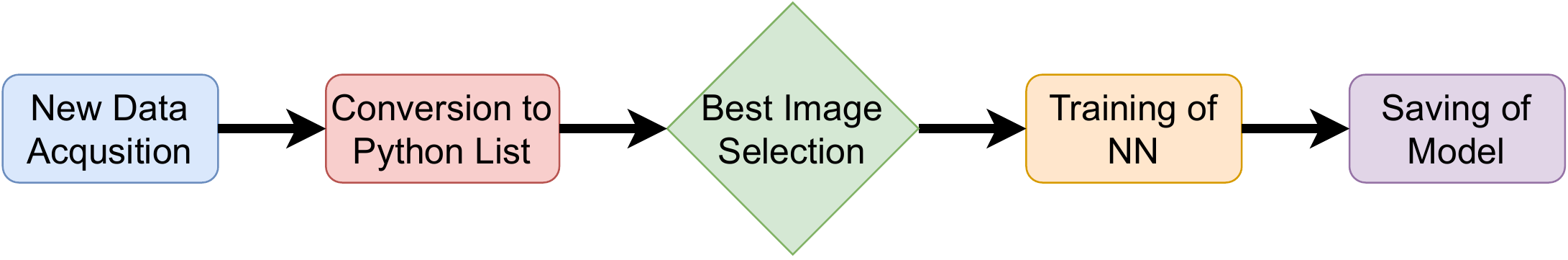}
\caption{The flow of Active Learning implementation. The data acquired using the Verasonics platform is shared with the NVIDIA CLARA AGX Kit. The data is converted to a python list and subsequently displayed to the user. The U-Net model is trained and saved with the user selected best image as the ground truth.} \label{fig4}
\end{figure}

The workflow starts with data acquisition using the Verasonics research US platform. Simultaneously, the acquired data along with the DAS, F-DMAS, MVDR, GCF images and the trained model image (except for the initial few cases) are shared with the NVIDIA CLARA AGX Developer Kit. In the NVIDIA CLARA AGX Developer kit, a python list of datasets are generated and the user is provided with the anonymized beamformed images through a streamlit \cite{streamlit} graphical user interface (GUI) as in Fig. \ref{fig3} to avoid bias in image selection. Once the user selects the best image according to his/her perception, the same is used to train the NN. The process is repeated for every new batch of data and on successful completion, the model is saved as the final trained model. The complete workflow is presented in Fig. \ref{fig4}.

The choice of the best image in ultrasound imaging is always challenging due to operational and system variability. In order to have a uniformity across users, a criteria as given below is employed for the best image selection:
\begin{enumerate}
    \item Determine regions of high intensity and compare for axial/lateral resolution.
    \item Determine regions of homogeneous speckle and compare for speckle resolution.
    \item Determine regions of contrast difference (e.g. cyst) and compare for contrast resolution.
\end{enumerate}
\subsubsection{Neural Network Architecture}
A U-Net \cite{unet2015} based architecture is selected for the proposed approach due to its capability for global and local feature learning. The proposed U-Net architecture consists of a double convolution layer followed by three hidden layers of downsampling, three hidden layers of upsampling, and the final convolution layer. The input to the model is the delay compensated RF data of size $m \times n \times N_{ch}$ where $m$ and $n$ are the number of pixels in depth and lateral directions respectively ($2400 \times 128$) while $N_{ch}$ is the number of channels. The double convolution layer changes the number of channels from 128 to 64 while keeping the data size in depth and lateral directions unchanged. The layer consists of a convolution layer followed by batch normalization and an activation function. The batch normalization is used after every convolution to increase the stability during training. The anti-rectifier activation function avoids 'dying' nodes when dealing with the RF data unlike ReLU \cite{luijten2020adaptive}. The final convolution layer changes the number of channels from 64 back to 128 (the same size as the input delay compensated RF data). The output of the model are the apodization weights of size same as the input. The delay compensated RF data are weighted by the apodization weights generated by the model, summed over the channels, envelope extracted and log compressed to generate the final B-mode image (as shown in Fig. \ref{fig1}). 

\begin{figure}[!htb]
\includegraphics[width=\textwidth]{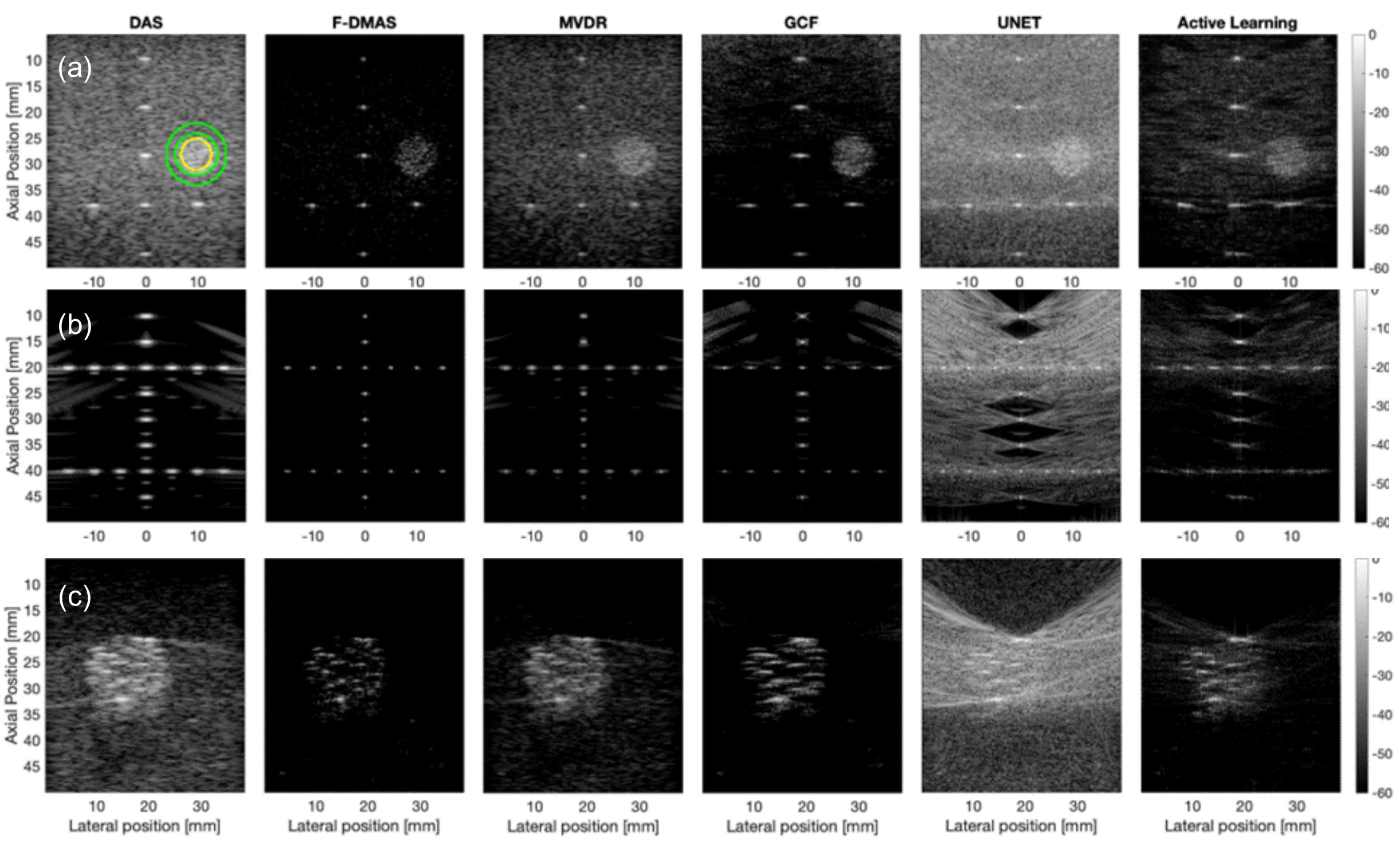}
\caption{Results from the \textit{in-silico} and \textit{in-vitro} experimental datasets (a) for a CIRS Multi-Purpose US Phantom (Model 040GSE). The region used for contrast measurement is indicated in the DAS image (b) PICMUS simulation dataset having point targets (c) An \textit{in-vitro} experimental dataset having inclusion targets (INS009 dataset of CUBDL and more details on the same is available in \cite{CUBDL2,CUBDL3}. The yellow circle and the green circle regions indicate the regions employed for CNR/CR calculations in Table \ref{table1}.} \label{fig5}
\end{figure}
\subsubsection{Training Details:}
The training data is collected from scans of the longitudinal and cross section scans of carotid artery, forearm muscles including brachioradialis and wrist of healthy volunteers following the principles outlined in the Helsinki Declaration of 1975, as revised in 2000. The reason for selection of these scans is to restrict to peripheral imaging ($< 5 cm$ depth) due to the planewave excitation. The models are implemented in Pytorch and trained on the 24GB RTX6000 GPU available with the NVIDIA CLARA AGX developer kit. The input for the model is the delay compensated RF data and the ground truth for the loss function is the user-selected beamformed image (one out of DAS, F-DMAS, MVDR, GCF, and the trained model (except for the initial few cases)). The mean squared error (MSE) between the U-Net and the user-selected beamformed images is the employed loss function. Considering the active learning scenario, the batch size is set to 1. The Adam optimizer is used with a learning rate of 1e-3 as in concurrence with \cite{mathews2021towards}. The trained models are tested on an \textit{in-silico} dataset with point targets and \textit{in-vitro} datasets provided by the plane-wave imaging challenge for medical ultrasound (PICMUS) \cite{PICMUS} and the challenge on ultrasound beamforming with deep learning (CUBDL) \cite{CUBDL1,CUBDL2,CUBDL3} of 2016 and 2020 IEEE International Ultrasonics Symposium, respectively. The training and test data have no overlap (i.e., the source of the data generated for the training and test are different).

\section{Results}
The section discusses the key results of the proposed approach in comparison with conventional and learning based beamforming approaches. The raw beamformed images without post-processing like speckle reduction or histogram corrections (thresholding, non-linear gamma correction, and other similar techniques for image enhancement) are presented to highlight the intrinsic differences between the beamformers. The quantitative comparisons are presented for contrast ratio (CR), contrast to noise ratio (CNR), axial and lateral full width half maximum (FWHM) metrics. The results are shown in Fig. \ref{fig5} and Tables \ref{table1} and \ref{table2}. The beamformed images with the conventional U-Net trained model with DAS as ground truth (i.e. non-user selected and typical NN approach) and with the active learning model are visually comparable to the images created with conventional beamforming algorithms. The results of active learning are certainly encouraging and are expected to improve with further training.

\renewcommand{\arraystretch}{1.25}
\setlength{\tabcolsep}{8pt}
\begin{table}[t]
\centering
\caption{Contrast and resolution metrics for DAS, F-DMAS, MVDR, GCF, U-NET (DAS), and the proposed active learning framework for Fig. \ref{fig5}(a).} \label{table1}
\resizebox{\textwidth}{!}{%
\begin{tabular}{c|c|c|c|c} 
\hline
    & \textbf{CR} & \textbf{CNR} & \textbf{Axial FWHM} & \textbf{Lateral FWHM} \\
                                &  & \textbf{(dB)} & \textbf{(mm)} & \textbf{(mm)} \\ \hline
\textbf{DAS \cite{GarciaDAS2021}}                    & 0.665       & 4.851             & 0.485                     & 1.327                      \\ \hline
\textbf{F-DMAS \cite{FDMAS2015}}                 & 0.677       & 4.614             & 0.458                    & 0.646                      \\ \hline
\textbf{MVDR \cite{MVDR2009}}                  & 0.669       & 1.736             & 0.481                    & 0.586                      \\ \hline
\textbf{GCF \cite{li2003adaptive}}                    & 0.623       & 6.677             & 0.433                    & 0.832                      \\ \hline
\textbf{UNET (DAS) \cite{mathews2021towards}, \cite{unet2015}}             & 0.714       & 2.947             & 0.488                    & 0.769                      \\ \hline
\textbf{Proposed UNET (Active Learning)} & 0.785       & 3.269             & 0.305                    & 0.650                      \\ 
\hline
\end{tabular}%
}
\end{table}
\setlength{\tabcolsep}{8pt}
\renewcommand{\arraystretch}{1.25}
\begin{table}[t]
\centering
\caption{Resolution metrics for DAS, F-DMAS, MVDR, GCF, UNET (DAS), and the proposed Active Learning framework for Fig. \ref{fig5}(b).}\label{table2}
\resizebox{\textwidth}{!}{%
\begin{tabular}{c|c|c}
\hline
                                & \textbf{Axial FWHM (mm)} & \textbf{Lateral FWHM (mm)} \\ \hline
\textbf{DAS \cite{GarciaDAS2021}}                    & 0.406                    & 1.131 \\ \hline
\textbf{F-DMAS \cite{FDMAS2015}}                 & 0.352                    & 0.485                      \\ \hline
\textbf{MVDR \cite{MVDR2009}}                   & 0.506                    & 0.519                      \\ \hline
\textbf{GCF \cite{li2003adaptive}}                    & 0.325                    & 0.377                      \\ \hline
\textbf{UNET (DAS) \cite{mathews2021towards}, \cite{unet2015}}             & 0.383                    & 0.399                      \\ \hline
\textbf{Proposed UNET (Active Learning)} & 0.297                    & 0.456                      \\ \hline
\end{tabular}%
}
\end{table}



\section{Discussion and Conclusions}
Recently, accelerating computationally intensive adaptive beamforming with NNs has been gaining momentum in US imaging. However, such models are static as they are trained against a particular algorithm (e.g. MVDR) by acquiring and storing large datasets (nearly 1000 diverse images) for training which is time and memory consuming. These concerns are overcome by the proposed active learning framework. The results from conventionally trained NN and the NN trained in the active learning framework are certainly encouraging. The results of the active learning framework is expected to improve with more training and tuning of the framework. From the statistics reported, it is estimated that the DAS beamforming has been selected in 22\%, the F-DMAS in 20\%, the MVDR and GCF in 29\% of the total number of datasets used for training. Using the same framework, the model can be shared among different institutes for US system agnostic training to make the model more robust under the federated learning framework. Also, the estimation of the image quality can be automated to accelerate the learning process independent of the user in a true active learning sense. 


\end{document}